\documentstyle[preprint,aps]{revtex}

\draft

\begin{document}
\title{Interplay of quantum magnetic and potential scattering around Zn or Ni
impurity ions in superconducting cuprates}
\author{Guang-Ming Zhang$^1$, Hui Hu$^2$, Lu Yu$^{2,3}$}
\address{$^1$Center for Advanced Study, Tsinghua University, Beijing 100084, China\\
$^2$Abdus Salam International Center for Theoretical Physics, P. O. Box 586,
Trieste 34100, Italy\\
$^3$Institute of Theoretical Physics and Interdisciplinary Center of
Theoretical Studies, \\
Academic Sinica, Beijing 100080, China}
\date{\today}
\maketitle

\begin{abstract}
To describe the scattering of superconducting quasiparticles from
non-magnetic (Zn) or magnetic (Ni) impurities in optimally doped high T$_c$
cuprates, we propose an effective Anderson model Hamiltonian of a localized
electron hybridizing with $d_{x^2-y^2}$-wave BCS type superconducting
quasiparticles with an attractive scalar potential at the impurity site. Due
to the strong local antiferromagnetic couplings between the original Cu ions
and their nearest neighbors, the localized electron in the Ni-doped
materials is assumed to be on the impurity sites, while in the Zn-doped
materials the localized electron is distributed over the four nearest
neighbor sites of the impurities with a dominant $d_{x^2-y^2}$ symmetric
form of the wave function. Since both scatterings from the localized
electron and the scalar potential are relevant, localized resonant states
due to their interplay are formed below the maximal superconducting gap.
With Ni impurities, two resonant states are formed above the Fermi level in
the local density of states at the impurity site, while for Zn impurities a
sharp resonant peak below the Fermi level dominates in the local density of
states at the Zn site, accompanied by a small and broad resonant state above
the Fermi level mainly induced by the potential scattering. This is exactly
what has been observed in the scanning tunneling microscopy experiments. In
both cases, there are no Kondo screening effects. From the calculated spin
relaxation functions, we find that the 3d localized electron in both Ni and
Zn doped materials displays a weak magnetic oscillation. This result is
consistent with the signal of a spin-1/2 magnetic moment exhibited by
nuclear magnetic resonance measurements in YBa$_2$Cu$_3$O$_{6+\delta }$
doped with Zn or Ni impurities. The local density of states and their
spatial distribution at the dominant resonant energy around the substituted
impurities are calculated for both cases, and they are in good agreement
with the experimental results of scanning tunneling microscopy in Bi$_2$Sr$_2
$CaCu$_2$O$_{8+\delta }$ with Zn or Ni impurities, respectively. Thus the
scanning tunneling and nuclear magnetic resonance experiments on Zn and Ni
substituted cuprates are interpreted self-consistently in a unified fashion.
\end{abstract}


\section{Introduction}

Recent experiments on $d$-wave superconducting cuprates have shown many
interesting phenomena in the presence of magnetic (Ni) and non-magnetic (Zn)
impurities substituting the Cu ions in the copper-oxygen plane. A series of
nuclear magnetic resonance (NMR) experiments on optimally doped YBa$_2$Cu$_3$%
O$_{6+\delta }$ with impurities \cite{nmr1,nmr2,nmr3} have shown that each
non-magnetic impurity (Zn) induces a local spin-1/2 magnetic moment on the
neighboring Cu sites, while a magnetic (Ni) impurity gives rise to a
localized spin-1/2 magnetic moment on the impurity site. On the other hand,
scanning tunneling microscopy (STM) has been used to probe the quasiparticle
scattering around single Zn or Ni impurity ions in optimally doped Bi$_2$Sr$%
_2$CaCu$_2$O$_{8+\delta }$ with high spatial and energy resolution \cite
{pan-nature,hudson-nature}. In the obtained STM spectra, a very sharp
localized resonance peak is found just below the zero-bias voltage on the Zn
impurity sites \cite{pan-nature}, while two localized resonant states are
formed well above the zero-bias voltage on the Ni impurity sites \cite
{hudson-nature}. The spatial dependence of the local density of states
(LDOS) in the vicinity of the individual impurity ions reveals the
characteristic features of $d_{x^2-y^2}$ wave superconducting ({\it d}SC)
states, but the pattern is different in these two cases.

A large number of theoretical papers have been devoted to the interpretation
of these beautiful experimental data. According to one school of thoughts,
this is due to a potential scattering, either spin-independent, or magnetic
(classical magnetic moment) \cite{balatsky,martin,flatte,tsuchiura,haas}. In
fact, the quasi-bound state around a potential scatter in $d$-wave
superconductors was discussed before the experiments were done \cite
{balatsky,flatte}. On the other hand, there is an alternative explanation
ascribing the resonant scattering as due to the Kondo effect in $d$-wave
superconductors \cite{sachdev,ting,vojta}, extending earlier studies of the
Kondo effect in systems with reduced density of states at the Fermi level 
\cite{fradkin,ingersent}. It is fair, however, to say that several important
issues remain unexplained, for example, the strong particle-hole asymmetry,
the difference between the Zn- and Ni-doped cases, concerning both spectral
distribution and spatial pattern. Thus a full understanding of these
experiments is still lacking.

In an earlier communication \cite{gmzhang}, we proposed an effective
Anderson Hamiltonian with hybridization of a localized electron with
quasiparticles in $d$-wave superconducting states to describe the scattering
on quantum magnetic impurities. In the limit of the Hubbard repulsion $%
U\rightarrow \infty $, we have shown the existence of a sharp resonance
above the Fermi level and absence of the Kondo screening. Moreover, a
``marginal Fermi liquid'' behavior was found for the impurity electron
self-energy in the strong coupling limit. In this paper we extend our study
to include the potential scattering effect on the impurity site. The
interplay of the quantum magnetic scattering and potential scattering turns
out to play a very important role in determining the spectral distribution
and spatial pattern.

The NMR \cite{nmr1,nmr2,nmr3} and neutron scattering \cite{sidis}
experiments show that Zn doping has a stronger disturbance to the
antiferromagnetic background in cuprates than Ni doping. There is a kind of
common understanding that Zn doping induces a local magnetic moment
distributed mostly over the nearest neighbor Cu ions, whereas Ni impurity
has an underscreened $S=1/2$ moment, sitting on the Ni ion itself. This
picture is consistent with the electron configurations. Zn$^{2+}$ ions have
a closed shell ($3d^{10}$, $S=0$), with an extra $d$-electron compared with
Cu$^{2+}$ which is not entirely sitting on the Zn site. The electronic
structure studies \cite{gupta} do show that the electron charge density is
higher on neighboring oxygen sites and nearest neighbor Cu$^{2+}$ ions. On
the other hand, Ni$^{2+}$ ions ($3d^8$, $S=1$) have an extra hole of spin $%
1/2$ on site, compared with Cu$^{2+}$ ions, and the deficit of charge is
concentrated on the Ni site.\cite{gupta} In this paper, we propose a unified
model to describe these two cases. We consider a local charge carrier ($3d$
electron in the Zn case and $3d$ hole in the Ni case, to be called Anderson
electrons for short, and such a treatment is allowed due to the
particle-hole symmetry of the BCS state) with a strong on-site Coulomb
repulsion $U$, is hybridizing with superconducting quasiparticles of $%
d_{x^2-y^2}$ symmetry. In the Ni-doped case, the Anderson electron is
sitting on the impurity site itself, whereas in the Zn-doped case, it is
distributed over the nearest neighbor Cu$^{2+}$ ions. The dominant
contribution comes from the linear combination with $d_{x^2-y^2}$ symmetry.
Moreover, we assume an attractive $\delta $-type potential scattering in
both cases, which can be justified to some extent within the three-band
Hubbard model .\cite{xiang}

In the above effective model, due to a strong interplay between the quantum
impurity scattering (described by the Anderson model) and the potential
scattering of the quasiparticles in $d$SC state, we find a sharp resonance
state below the Fermi energy mainly due to the Anderson electron, and a
broad potential scattering peak in the positive energy in the LDOS for the
Zn-doped case. In contrast, two resonance peaks due to Anderson electron and
potential scattering are formed well above the Fermi level for the Ni-doped
case. These distinct features fully agree with STM experiments \cite
{pan-nature,hudson-nature}. Moreover, the spatial pattern of the resonance
states differs in these two cases, also in agreement with experiments. The
calculated dynamical spin response functions show typical structures
corresponding to the existence of local magnetic moments. These results
demonstrate convincingly that the proposed unified model describes the
essential physics of these experiments self-consistently.

The rest of the paper is organized as follows. In section II, the nickel
impurity substitution is considered, with both Anderson electron and
potential scattering center located at the impurity site itself. In section
III, we discuss the interplay of quantum magnetic scattering on the Anderson
electron distributed over nearest neighbors and on-site potential
scattering, corresponding to the Zn-doped case. Finally, concluding remarks
are made in section IV.

\section{ Model Calculation for Nickel Impurity Substitution}

To construct a simple model, we assume that a BCS-type weak coupling theory
is applicable as a phenomenological framework for high-T$_c$ optimally doped
superconductors, though the underlying mechanisms are very different. When
the Cu ions on the CuO$_2$ plane are replaced by magnetic Ni impurities,
there will be one extra $3$d hole on each impurity ion. Due to the
particle-hole symmetry of the {\it d}SC quasiparticles, we can work in the
electron representation, and assume that the magnetic Ni impurity is
effectively described by the Anderson localized electron with a strong
Hubbard repulsion. Moreover, a scalar {\it attractive} potential at the
impurity site has been added to the model Hamiltonian as caused by
interactions with the nearest neighbor Cu ions. When the correlations
between the magnetic impurities on different sites are ignored, the model
Hamiltonian is defined as 
\begin{eqnarray}
{\cal H} &=&\sum\limits_{{\bf k}\sigma }\epsilon _{{\bf k}}c_{{\bf k}\sigma
}^{+}c_{{\bf k}\sigma }+\sum\limits_{{\bf k}}\Delta _{{\bf k}}\left( c_{{\bf %
k}\uparrow }^{+}c_{-{\bf k}\downarrow }^{+}+h.c.\right)  \nonumber \\
&&+\frac WN\sum\limits_{{\bf k,k^{\prime }},\sigma }c_{{\bf k}\sigma }^{+}c_{%
{\bf k^{\prime }}\sigma }+\epsilon _d\sum\limits_\sigma d_\sigma ^{+}d_\sigma
\nonumber \\
&&+\frac V{\sqrt{N}}\sum\limits_{{\bf k}\sigma }\left( c_{{\bf k}\sigma
}^{+}d_\sigma +h.c.\right) +Ud_{\uparrow }^{+}d_{\uparrow }d_{\downarrow
}^{+}d_{\downarrow },
\end{eqnarray}
where $\epsilon _{{\bf k}}=\hbar ^2k^2/(2m)-\epsilon _F$ is the dispersion
of the normal state electrons, $\Delta _{{\bf k}}=\Delta _0\cos 2\theta $ is
the {\it d}SC order parameter with the gap amplitude $\Delta _0$, and $W<0$
is the strength of the attractive potential. When Nambu spinors are
introduced 
\[
\hat{\psi}_{{\bf k}}=\left( 
\begin{array}{c}
c_{{\bf k}\uparrow } \\ 
c_{-{\bf k}\downarrow }^{+}
\end{array}
\right) ,\qquad \widehat{\varphi }=\left( 
\begin{array}{c}
d_{\uparrow } \\ 
d_{\downarrow }^{+}
\end{array}
\right) , 
\]
we rewrite the model Hamiltonian in a matrix from 
\begin{eqnarray}
{\cal H} &=&\sum\limits_{{\bf k}}\hat{\psi}_{{\bf k}}^{+}\left( \epsilon _{%
{\bf k}}\sigma _z+\Delta _{{\bf k}}\sigma _x\right) \hat{\psi}_{{\bf k}}+%
\frac WN\sum\limits_{{\bf k,k^{\prime }}}\hat{\psi}_{{\bf k}}^{+}\sigma _z%
\hat{\psi}_{{\bf k^{\prime }}}  \nonumber \\
&&+\frac V{\sqrt{N}}\sum\limits_{{\bf k}}\left( \hat{\psi}_{{\bf k}%
}^{+}\sigma _z\widehat{\varphi }+h.c.\right)  \nonumber \\
&&+(\epsilon _d+\frac U2)(\widehat{\varphi }^{+}\sigma _z\widehat{\varphi }%
+1)-\frac U2(\widehat{\varphi }^{+}\widehat{\varphi }-1)^2,
\end{eqnarray}
where $\sigma _z$ and $\sigma _x$ are Pauli matrices. The equations of
motion for the {\it d}SC quasiparticles yield the generalized {\bf T}-matrix 
{\it exactly} 
\begin{eqnarray}
\hat{T}(i\omega _n) &=&V^2[\sigma _z-W\hat{G}_0(i\omega _n)]^{-1}\hat{G}%
_d\left( i\omega _n\right) [\sigma _z-W\hat{G}_0(i\omega _n)]^{-1}  \nonumber
\\
&&+W[\sigma _z-W\hat{G}_0(i\omega _n)]^{-1},
\end{eqnarray}
where $\hat{G}_0(i\omega _n)=\frac 1N\sum_{{\bf k}}\left( i\omega
_n-\epsilon _{{\bf k}}\sigma _z-\Delta _{{\bf k}}\sigma _x\right) ^{-1}$ is
the GF of the {\it d}SC quasiparticles at the impurity site. When the scalar
potential is absent, the T-matrix is reduced to 
\begin{equation}
\hat{T}\left( i\omega _n\right) =V^2\sigma _z\hat{G}_d\left( i\omega
_n\right) \sigma _z,
\end{equation}
which was obtained previously in the single impurity Anderson model in $d$SC
electron state \cite{gmzhang}. On the other hand, if only the scalar
potential is present, it becomes 
\begin{equation}
\hat{T}\left( i\omega _n\right) =[(W\sigma _z)^{-1}-\hat{G}_0(i\omega
_n)]^{-1},
\end{equation}
which is the same result as derived by Balatsky {\it et al}. \cite{balatsky}%
. From the generalized T-matrix, one can clearly see that {\it both
hybridization with the localized electron and potential scattering are
equally important in determining the low energy excitations}. At zero
temperature, analytical continuation is used to calculate the perturbed GF
through the following relation: 
\[
\hat{G}\left( {\bf r},{\bf r}^{\prime };\omega \right) =\hat{G}^0\left( {\bf %
r}-{\bf r}^{\prime },\omega \right) +\hat{G}^0\left( {\bf r},\omega \right) 
\hat{T}\left( \omega \right) \hat{G}^0\left( -{\bf r}^{\prime },\omega
\right) . 
\]
The LDOS of the {\it d}SC quasiparticles around the impurity is given by 
\[
N\left( {\bf r},\omega \right) =-\frac 1\pi 
\mathop{\rm Im}%
\hat{G}_{11}\left( {\bf r},{\bf r};\omega \right) . 
\]
We would emphasize that the above relation between the GF of the {\it d}SC
quasiparticles and the Anderson electron is {\it exact}, and the only
quantity to determine is $\hat{G}_d(\omega )$, which will be calculated
approximately.

When we take the infinite $U$ limit, the localized electron operator is
expressed as $\widehat{\varphi }^{+}=\left( 
\begin{array}{cc}
f_{\uparrow }^{+}b, & f_{\downarrow }b^{+}
\end{array}
\right) $ in the slave-boson representation \cite{barnes,coleman}, where the
fermion $f_\sigma $ and the boson $b$ describe the singly occupied and hole
states, respectively. The constraint $b^{+}b+\sum\limits_\sigma f_\sigma
^{+}f_\sigma =1$ has to be imposed. When a mean field approximation is
applied, the boson operators $b$ and $b^{+}$ are replaced by a c-number $b_0$%
, and the constraint is treated by introducing a chemical potential $\lambda
_0$. Therefore, the mean field Hamiltonian is written as 
\begin{eqnarray}
{\cal H}_{mf} &=&\sum\limits_{{\bf k}}\hat{\psi}_{{\bf k}}^{+}\left(
\epsilon _{{\bf k}}\sigma _z+\Delta _{{\bf k}}\sigma _x\right) \hat{\psi}_{%
{\bf k}}+\frac WN\sum\limits_{{\bf k,k^{\prime }}}\hat{\psi}_{{\bf k}%
}^{+}\sigma _z\hat{\psi}_{{\bf k^{\prime }}}  \nonumber \\
&&+\frac{\tilde{V}}{\sqrt{N}}\sum\limits_{{\bf k}}\left( \hat{\psi}_{{\bf k}%
}^{+}\sigma _z\widehat{\phi }+h.c.\right) +\tilde{\epsilon}_d\widehat{\phi }%
^{+}\sigma _z\widehat{\phi }  \nonumber \\
&&+\tilde{\epsilon}_d+\lambda _0(b_0^2-1).
\end{eqnarray}
where $\widehat{\phi }^{+}=\left( 
\begin{array}{cc}
f_{\uparrow }^{+}, & f_{\downarrow }
\end{array}
\right) $denotes the Nambu spinors of the localized electron and the
renormalized parameters $\tilde{\epsilon}_d=\epsilon _d+\lambda _0$ and $%
\tilde{V}=b_0V$. It has been known that the slave-boson mean field
approximation produces the correct low energy physics of the conventional
Anderson impurity model in the entire parameter range at zero temperature,
though strong quantum fluctuations are present near the Fermi surface from
the gapless excitations of the conduction electrons \cite{coleman2}. In the
present model, we are dealing with $d$SC quasiparticles, where a
superconducting gap exists on most part of the Fermi surface except four
nodal points, we thus expect that the slave-boson mean field approximation
should work better than the conventional case, giving rise to more reliable
results at zero temperature.

The standard techniques lead to $\hat{G}_f(i\omega _n)=\left[ i\omega _n-%
\tilde{\epsilon}_d\sigma _z-\hat{\Sigma}_f\left( i\omega _n\right) \right]
^{-1}$, where the self-energy of the localized electron becomes, 
\begin{equation}
\hat{\Sigma}_f\left( i\omega _n\right) =\tilde{V}^2[\hat{G}_0^{-1}(i\omega
_n)-W\sigma _z]^{-1}.
\end{equation}
At $T=0$, the ground-state energy changes due to the presence of localized
electron and the scalar potential are:

\begin{eqnarray*}
\delta \varepsilon &=&-\frac 1\pi \int\limits_0^Dd\omega \ln \left[ \omega
^2\left( 1+b_0^2\beta \right) ^2+\left[ \tilde{\epsilon}_d-b_0^2(W/V^2)%
\omega ^2\alpha \beta \right] ^2\right] \\
&&+\tilde{\epsilon}_d+\lambda _0\left( b_0^2-1\right) ,
\end{eqnarray*}
where $D$ is the band width of the normal conduction electrons, $\alpha
(\omega )=\frac 1N\sum\limits_{{\bf k}}\frac{V^2}{\omega ^2+\epsilon _{{\bf k%
}}^2+\Delta _{{\bf k}}^2}$, and $\beta (\omega )=\frac{\alpha (\omega )}{%
1+(W/V^2)\omega ^2\alpha ^2(\omega )}$. The corresponding saddle-point
equations are derived as 
\begin{eqnarray}
\lambda _0 &=&\frac 1\pi \int\limits_0^Dd\omega \frac{2\omega ^2\beta
(1+b_0^2\beta )-2(W/V^2)\omega ^2\alpha \beta \left[ \tilde{\epsilon}%
_d-b_0^2(W/V^2)\omega ^2\alpha \beta \right] }{\omega ^2\left( 1+b_0^2\beta
\right) ^2+\left[ \tilde{\epsilon}_d-b_0^2(W/V^2)\omega ^2\alpha \beta
\right] ^2}, \\
b_0^2 &=&\frac 1\pi \int\limits_0^Dd\omega \frac{2\left[ \tilde{\epsilon}%
_d-b_0^2(W/V^2)\omega ^2\alpha \beta \right] }{\omega ^2\left( 1+b_0^2\beta
\right) ^2+\left[ \tilde{\epsilon}_d-b_0^2(W/V^2)\omega ^2\alpha \beta
\right] ^2}.
\end{eqnarray}
For given parameters $D$, $\Delta _0$, $\Gamma =\pi N_FV^2$, $\epsilon _d$,
and $c=\pi N_F|W|$, one can obtain saddle point values of $b_0$ and $\lambda
_0$. In the following calculations, we choose $\Delta _0$ as the energy
unit, $D/\Delta _0=20$, and $\Gamma /\Delta _0=0.2$.

The phase diagram can be obtained for a specific value of the potential
scattering strength $c$, similar to the phase diagram given previously \cite
{gmzhang}. When $\epsilon _d$ is less than a threshold value, $b_0^2$ is
zero, leading to a free local magnetic moment decoupled from the {\it d}SC
quasiparticles. Above the threshold value of $\epsilon _d$, $b_0^2$ rises
sharply and then saturates. The low energy behavior of the model is
controlled by the nearly empty orbital state, characterized by a strong
hybridization between the Anderson electron and the {\it d}SC
quasiparticles. In the hole representation, the empty orbital phase
corresponds to a single hole occupation at the Ni impurity site. Such a
situation can not be realized in the conventional Anderson impurity model,
where the low energy behavior is always controlled by the Kondo screening
fixed point in the single electron occupied phase. The {\it d}SC
quasiparticles actually provide us with such a possibility to explore new
low-energy physics in the empty orbital regime --- the strong coupling
behavior in the single hole occupied regime. In a sense, such a strong
coupling behavior is also universal, corresponding to a new strong coupling
fixed point without any characteristic energy scale. However, the scalar
potential will introduce some extra features to the LDOS around the
impurities.

We have seen that both scalar potential and localized electron scatterings
are relevant in the {\it d}SC state, and two resonant states are formed with
a dominant sharp localized resonance {\it above} the Fermi level. In Fig. 1,
we plot the DOS for the localized electron $N_d\left( \omega \right) $ with
different values of $\epsilon _d$ for two potential scattering strengths $c=1
$ and $2$. For a small $\epsilon _d$, the resonant peak positions can be
roughly determined as 
\begin{equation}
\Omega _{res}^1\approx \tilde{\epsilon}_d\left[ 1-b_0^2\left( \frac{2\Gamma 
}{\pi \Delta _0}\right) \ln \frac{4\Delta _0}{\tilde{\epsilon}_d}\right] ,%
\text{ }\Omega _{res}^2\simeq \Delta _0\frac \pi {2c\ln (8c/\pi )}.
\end{equation}
Clearly, the former resonance peak is mainly induced by the localized
Anderson electron, while the latter weaker resonance is caused by the
attractive potential. As $\epsilon _d$ moves up, the magnitude of the first
localized resonance $\Omega _{res}^1$ decreases and becomes broader with
shifting upward of the peak position, while the magnitude of the second
resonance $\Omega _{res}^2$ emerges as a pronounced peak at a relative high
energy. When $\epsilon _d\sim 0.30$, the profile of the two consecutive
resonances agrees well with the measured DOS at individual Ni site \cite
{hudson-nature}. This does also confirm our assumption of the {\it %
attractive }scalar potential at the Ni impurity site, because the
corresponding localized resonant state would appear below the Fermi energy
if the scalar potential scattering is repulsive.

The nature of the double resonant peaks in the DOS of localized electrons
can also be understood from the imaginary part of its retarded self-energy:

\begin{equation}
\mathop{\rm Im}%
\Sigma _{f,}^{1,1}\left( \omega +i0^{+}\right) =%
\mathop{\rm Im}%
\frac{b_0^2\Gamma }{(\pi N_F)G_0^{-1}(\omega +i0^{+})+c},
\label{self-energy}
\end{equation}
which has been plotted in Fig.2a. In contrast to the case in the absence of
attractive scalar potential (solid line), a single peak appears above the 
Fermi energy (dashed line), corresponding to the pole of Eq. (\ref
{self-energy}). This peak becomes narrower with increasing the potential
strength $c$ (dash-dotted line).  Moreover, the existence of the scalar
potential greatly suppresses the spectrum weight of the self-energy far away
from the Fermi level. However, the marginal Fermi liquid behavior \cite
{varma} obtained previously is preserved: $%
\mathop{\rm Im}%
\Sigma _{f,}^{1,1}\left( \omega \right) \sim -b_0^2\Gamma \left( \frac{%
|\omega |}{\Delta _0}\right) $as $\omega \rightarrow 0$.

Furthermore, in order to reveal the characteristic behavior in the regime of
the empty orbital (single hole) localized state, we calculate the dynamic
spin response function of the localized electron 
\begin{equation}
\chi _d(t,t^{\prime })=\langle \langle S_d^z(t):S_d^z(t^{\prime })\rangle
\rangle
\end{equation}
corresponding to particle-hole collective excitations, from which the spin
relaxation function $S_d(\omega )=-\frac 1\pi \frac{\chi _d^{\prime \prime
}(\omega +i0^{+})}\omega $ can be evaluated. $S_d(\omega )$ is related to
the neutron scattering cross-section and the calculated value  is displayed
in Fig.2b. There is a pair of symmetric peaks demonstrating {\it a weak
magnetic oscillation} for the $3d$ localized hole, in contrast to the
universal result of one central resonant peak obtained in the Kondo regime
of the conventional Anderson impurity model \cite{costi}. Such a behavior is
consistent with a local spin 1/2 magnetic moment on the Ni impurity site
observed in the NMR measurements \cite{nmr3}. We also notice that closer the
localized resonant state is to the Fermi level, stronger magnetic
oscillation the localized hole exhibits. This can be understood by
considering the linear energy dependence of the LDOS for the {\it d}SC
quasiparticles: the localized resonant state becomes more and more
delocalized as one moves away from the Fermi level. As for the self-energy,
the presence of potential scattering ($c\neq 0$) reduces substantially the
magnitude of the spin relaxation function.

Now consider the LDOS of the {\it d}SC quasiparticles $N\left( {\bf r}%
,\omega \right) $ in the spatial range $0<{\bf r\leq }$ $\xi $. In high T$_c$
{\it d}SC state, $\xi $ is about $10{\rm \AA }$, or roughly 3 lattice
spacings. In Fig. 3, the LDOS vs frequency is shown for $r{\bf =}0.07\xi $
from the impurity site along the vertical (or horizontal) and diagonal
directions in real space. Similar to the LDOS for the localized electron,
there are two resonances above the Fermi energy along the diagonal (the gap
nodal) directions and the LDOS is entirely hole-like. In contrast, for
vertical and horizontal  (the gap maximum) directions, two more resonant
peaks appear below the Fermi level as well, which are slightly asymmetric in
the line shape and strongly reminiscent of the localized bound states
induced by paramagnetic impurities in the conventional $s$-wave
superconducting state \cite{yulu}. In both of these cases, increasing the
potential strength will sharpen and enhance the localized resonances
significantly. As the impurity energy level $\epsilon _d $ moves up, the
quasiparticle resonances become broader, exhibiting a similar dependence as
the resonance of the Anderson electron itself.

It is very important to calculate the spatial distribution of the LDOS of 
{\it d}SC quasiparticles to compare with corresponding STM results. The LDOS
around the impurity at the first resonance energy is displayed in Fig.4a\
for the hole-like excitations $\omega =-\Omega _{res}^1$ and in Fig.4b for
the electron-like excitations $\omega =\Omega _{res}^1$ as a function of
spatial variables for $c=1,2,$ and $3$ with $\epsilon _d/\Delta _0=-0.05$ in
a logarithmic intensity scale. The quasiparticle resonances induced by
substituted impurities are highly localized around the impurity, and the
spatial oscillation of these resonant states is visible. The largest
amplitude of the quasiparticle resonance occurs in the neighborhood of the
impurity, and the local electronic structures distinctly differ in Fig.4a
from that in Fig.4b. For $\omega =\Omega _{res}^1$, the LDOS exhibits a
four-fold symmetry along the directions of the gap nodes for all distances,
consistent with the {\it d}SC quasiparticles. For $\omega =-\Omega _{res}^1$%
, the LDOS is strongly enhanced in the gap maxima directions at distances $%
r\ll \xi $. Further away from the impurity ($r{\bf \sim }\xi $), it is
confined to the neighborhood of the diagonal directions, preserving the
four-fold symmetry. At the same time, the LDOS of the {\it d}SC
quasiparticles at the impurity site (the center of the diagram) and around
the impurity site get enhanced by increasing the strength of the scalar
potential. In fact, these features have been seen in the STM experiment \cite
{hudson-nature}, namely, the resonance at positive energy is enhanced in the
nodal directions, while the resonance at negative energy is more visible in
the anti-nodal directions. To the best of our knowledge, these distinctive
features have not been explained by other theoretical models considered so
far. The absence of particle-hole symmetry is natural  in our consideration,
due to the quantum origin of the resonant states, and it is in full
agreement with STM experiments.\cite{hudson-nature}

\section{Model Calculation for Zinc Impurity Substitution}

In the high T$_c$ cuprates, the substitution of Cu ions by Zn impurities
adds an additional $3d$ electron to each Cu ion. By carefully analyzing the
experimental results \cite{nmr1,nmr2,pan-nature}, one can notice that local
antiferromagnetic correlations around the non-magnetic impurities get
enhanced. As a result, there should be a localized magnetic moment
distributed mainly over the four nearest neighbors of Zn impurities with an 
{\it attractive} scalar potential at the impurity. In the following, we will
work in the electron representation, and assume that the localized state is
still described by the Anderson localized electron with a strong on-site
Hubbard repulsion. The effective model Hamiltonian is thus given by: 
\begin{eqnarray}
{\cal H} &=&\sum\limits_{{\bf k}\sigma }\epsilon _{{\bf k}}c_{{\bf k}\sigma
}^{+}c_{{\bf k}\sigma }+\sum\limits_{{\bf k}}\Delta _{{\bf k}}\left( c_{{\bf %
k}\uparrow }^{+}c_{-{\bf k}\downarrow }^{+}+h.c.\right)  \nonumber \\
&&+\frac WN\sum\limits_{{\bf k,k^{\prime }},\sigma }c_{{\bf k}\sigma }^{+}c_{%
{\bf k^{\prime }}\sigma }+\epsilon _d\sum\limits_\sigma d_\sigma ^{+}d_\sigma
\nonumber \\
&&+\frac 1{\sqrt{N}}\sum\limits_{{\bf k}\sigma }V_{{\bf k}}\left( c_{{\bf k}%
\sigma }^{+}d_\sigma +h.c.\right) +Ud_{\uparrow }^{+}d_{\uparrow
}d_{\downarrow }^{+}d_{\downarrow }.
\end{eqnarray}
Compared with the previous model Hamiltonian for the nickel substitution,
the momentum dependence of the hybridization strength takes the form of $V_{%
{\bf k}}=V\cos 2\theta $, because the Anderson electron sitting on the four
nearest neighbor sites of the Zn impurity hybridizes with the {\it d}SC
quasiparticles with a dominant $d_{x^2-y^2}$ symmetric form of the wave
function. This is the most effective coupling channel in the {\it d}SC \cite
{sachdev,vojta}. The model Hamiltonian now describes an {\it extended}
Anderson quantum ``impurity'' coupled with the {\it d}SC quasiparticles. In
terms of the Nambu spinors, the GF of the conduction electrons is deduced
from the equations of motion 
\begin{equation}
\hat{G}({\bf k,k}^{{\bf \prime }};i\omega _n)=\delta _{{\bf k,k}^{\prime }}%
\hat{G}_0({\bf k},i\omega _n)+\frac 1N\hat{G}_0({\bf k},i\omega _n)\hat{T}_{%
{\bf k,k}^{\prime }}(i\omega _n)\hat{G}_0({\bf k}^{\prime },i\omega _n),
\end{equation}
where the generalized {\bf T}-matrix becomes momentum dependent. After
Fourier transformation, the above equation is expressed explicitly as

\begin{eqnarray}
&&\hat{G}\left( {\bf r},{\bf r}^{\prime };i\omega _n\right)  \nonumber \\
&=&\hat{G}^0\left( {\bf r}-{\bf r}^{\prime },i\omega _n\right) +\hat{G}%
^0\left( {\bf r},i\omega _n\right) \hat{T}_0\left( i\omega _n\right) \hat{G}%
^0\left( -{\bf r}^{\prime },i\omega _n\right)  \nonumber \\
&&+V^2\hat{G}_i^0\left( {\bf r},i\omega _n\right) \sigma _z\hat{G}_d\left(
i\omega _n\right) \sigma _z\hat{G}_i^0\left( -{\bf r}^{\prime },i\omega
_n\right)  \nonumber \\
&&+V^2\hat{G}_i^0\left( {\bf r},i\omega _n\right) \sigma _z\hat{G}_d\left(
i\omega _n\right) \sigma _z\hat{G}^0\left( {\bf r}_0,i\omega _n\right) \hat{G%
}_p^0\left( i\omega _n\right) \hat{G}^0\left( -{\bf r}^{\prime },i\omega
_n\right)  \nonumber \\
&&+V^2\hat{G}^0\left( {\bf r},i\omega _n\right) \hat{G}_p^0\left( i\omega
_n\right) \hat{G}^0\left( {\bf r}_0,i\omega _n\right) \sigma _z\hat{G}%
_d\left( i\omega _n\right) \sigma _z\hat{G}_i^0\left( -{\bf r}^{\prime
},i\omega _n\right) ,
\end{eqnarray}
with a momentum independent part of the T-matrix 
\[
\hat{T}_0\left( i\omega _n\right) =\hat{G}_p^0\left( i\omega _n\right) +V^2%
\hat{G}_p^0\left( i\omega _n\right) \hat{G}^0\left( {\bf r}_0,i\omega
_n\right) \sigma _z\hat{G}_d\left( i\omega _n\right) \sigma _z\hat{G}%
^0\left( {\bf r}_0,i\omega _n\right) \hat{G}_p^0\left( i\omega _n\right) , 
\]
where $\hat{G}_p^0\left( i\omega _n\right) =W[\sigma _z-W\hat{G}_0(i\omega
_n)]^{-1}$ is the resulting GF for the {\it d}SC quasiparticles at the
impurity site in the presence of the potential scattering only, $\hat{G}%
^0\left( {\bf r}_0,i\omega _n\right) =\frac 1{NV}\sum_{{\bf k}}V_{{\bf k}}%
\hat{G}_0({\bf k},i\omega _n)$ corresponds to the GF at the localized
electron site, and $\hat{G}_i^0\left( {\bf r},i\omega _n\right) =\frac 1{NV}%
\sum_{{\bf k}}V_{{\bf k}}\hat{G}_0({\bf k},i\omega _n)e^{i{\bf k\cdot r}}$
is the corresponding Fourier transform of the quasiparticle GF with a form
factor of the localized electron. In the present continuum model, the
positions of the nearest neighbor of the impurity ${\bf r}_0$ will be set
according to our chosen parameters.

When the slave-boson mean field approximation is applied ,  the
quasiparticle GF for the localized electron is derived as $\hat{G}_f(i\omega
_n)=\left[ i\omega _n-\tilde{\epsilon}_d\sigma _z-\hat{\Sigma}_f\left(
i\omega _n\right) \right] ^{-1}$, where the corresponding self-energy is
given by 
\begin{equation}
\hat{\Sigma}_f\left( i\omega _n\right) =\tilde{V}^2\sigma _z[\hat{G}_s^0(%
{\bf r}_0,i\omega _n)+\hat{G}^0\left( {\bf r}_0,i\omega _n\right) \hat{G}%
_p^0\left( i\omega _n\right) \hat{G}^0\left( {\bf r}_0,i\omega _n\right)
]^{-1}\sigma _z,
\end{equation}
where $\hat{G}_s^0({\bf r}_0,i\omega _n)=\frac 1{NV^2}\sum_{{\bf k}}V_{{\bf k%
}}^2\hat{G}_0({\bf k},i\omega _n)$. In order to carry out further
calculations, the various GFs introduced above are simplified as follows 
\begin{eqnarray}
\hat{G}_0(i\omega _n) &=&-i\omega _n\left( \frac{\pi N_F}\Gamma \right)
\alpha (i\omega _n),  \nonumber \\
\hat{G}^0\left( {\bf r}_0,i\omega _n\right) &=&-\Delta _0\left( \frac{\pi N_F%
}\Gamma \right) \gamma (i\omega _n)\sigma _x,  \nonumber \\
\hat{G}_s^0({\bf r}_0,i\omega _n) &=&-i\omega _n\left( \frac{\pi N_F}\Gamma
\right) \gamma (i\omega _n),
\end{eqnarray}
with 
\begin{eqnarray}
\alpha (\omega ) &=&\frac{2\Gamma }{\pi \Delta _0}\frac 1{\sqrt{1+\left(
\omega /\Delta _0\right) ^2}}K\left( \frac 1{\sqrt{1+\left( \omega /\Delta
_0\right) ^2}}\right) ; \\
\gamma (\omega ) &=&\frac{2\Gamma }{\pi \Delta _0}\left[ \sqrt{1+\left(
\omega /\Delta _0\right) ^2}E\left( \frac 1{\sqrt{1+\left( \omega /\Delta
_0\right) ^2}}\right) -\frac{\left( \omega /\Delta _0\right) ^2}{\sqrt{%
1+\left( \omega /\Delta _0\right) ^2}}K\left( \frac 1{\sqrt{1+\left( \omega
/\Delta _0\right) ^2}}\right) \right] ,
\end{eqnarray}
where $K(x)$ and $E(x)$ are the complete elliptical integrals of the first
and second kind, respectively. At zero temperature, the ground-state energy
changes due to the presence of Zn impurity and scalar potential are:

\begin{eqnarray*}
\delta \varepsilon  &=&-\frac 1\pi \int\limits_0^Dd\omega \ln \left( \omega
^2\left[ 1+b_0^2\gamma (\omega )A_1(\omega )\right] ^2+\left[ \tilde{\epsilon%
}_d-b_0^2A_2(\omega )\right] ^2\right)  \\
&&+\tilde{\epsilon}_d+\lambda _0\left( b_0^2-1\right) ,
\end{eqnarray*}
where $D$ is the band width of the normal conduction electrons, and 
\begin{eqnarray}
A_1(\omega ) &=&1+\frac{\left( c\Delta _0\right) ^2\alpha (\omega )\gamma
(\omega )}{\Gamma ^2+c^2\omega ^2\alpha ^2(\omega )},  \nonumber \\
A_2(\omega ) &=&\frac{-c\Gamma \Delta _0^2\text{ }\gamma ^2(\omega )}{\Gamma
^2+c^2\omega ^2\alpha ^2(\omega )}.
\end{eqnarray}
The saddle-point equations are thus derived as 
\begin{eqnarray}
\lambda _0 &=&\frac 1\pi \int\limits_0^Dd\omega \frac{2\omega ^2\gamma
A_1(1+b_0^2\gamma A_1)-2A_2(\tilde{\epsilon}_d-b_0^2A_2)}{\omega ^2\left[
1+b_0^2\gamma (\omega )A_1(\omega )\right] ^2+\left[ \tilde{\epsilon}%
_d-b_0^2A_2(\omega )\right] ^2}, \\
b_0^2 &=&\frac 1\pi \int\limits_0^Dd\omega \frac{2\left( \tilde{\epsilon}%
_d-b_0^2A_2\right) }{\omega ^2\left[ 1+b_0^2\gamma (\omega )A_1(\omega
)\right] ^2+\left[ \tilde{\epsilon}_d-b_0^2A_2(\omega )\right] ^2}.
\end{eqnarray}

Notations here are the same as used in Eqs.(8) and (9). These equations are
solved self-consistently to obtain the phase diagram which is essentially
the same as for the Ni-doped case, {\it i.e.} the physics is dominated by
the single-hole occupied state. Again, due to the presence of both
attractive scalar potential and quantum impurity scatterings, localized
resonant states in the quasiparticle spectrum are induced. However, unlike
the Ni-doped case, a dominant sharp resonance associated with the localized
Anderson electron is formed above the Fermi level, while a small and broad
resonance state associated with the scalar potential appears below the Fermi
level. This difference is mainly due to the different location of the
Anderson electron: on-site vs nearest neighbors. We plot the DOS for the
localized electron $N_d\left( \omega \right) $ with different values of $%
\epsilon _d$ for two potential scattering strengths $c=2$ and $5$ in Fig.5.
As $\epsilon _d$ moves up, the magnitude of the resonance above the Fermi
level slightly decreases and the peak position is shifted away from the
Fermi energy.

To compare with results for Ni substitution, the imaginary part of the
retarded self-energy for the localized electron is calculated for the
strengths of the potential scattering $c=1,2,5$ and $\epsilon _d/\Delta
_0=-0.2$, displayed in Fig.6a. The main difference from the Ni doping case
is that a single narrow resonance peak shows up below the Fermi level and
becomes sharper as the strength of potential scattering is increasing. In
Fig.6b, the spin relaxation function for the localized electron $S_d(\omega
) $ is also shown. A magnetic oscillation is displayed and the amplitude is 
{\it one order of magnitude larger } than that in the Ni case (Please note
the solid line corresponds to $c=1$ now), implying the single $3d$ hole on
the nearest neighbors of the Zn impurity atom does display a substantial
local magnetic moment! \cite{nmr1,nmr2} Certainly, we do not claim here a
full explanation for the local magnetic moment formation induced by Zn
impurity doping. However, our calculations do indicate the existence of a
local magnetic moment on the nearest neighbors of the Zn impurity atom.

To confront the STM\ experimental results, we calculate the LDOS of the {\it %
d}SC quasiparticles around the Zn impurity sites under $\epsilon _d/\Delta
_0=-0.3,-0.1$ for two strengths of scalar potential $c=2,4 $, displayed in
Fig.7. The LDOS on the Zn impurity site has clearly shown that there are a
very sharp resonance just below the Fermi level and a small and broad
resonance above the Fermi level. The former is induced by the hybridization
between the localized hole with the {\it d}SC quasiparticles, while the
latter is caused by the potential scattering. Moreover, in contrast to the
Ni substitution result, the DOS near the gap edges are completely suppressed
to zero, which is consistent with the fact that the superconductivity may be
suppressed by Zn impurities stronger than that by the Ni impurities. On the
nearest neighbor sites ${\bf r}_0=(\pm a,0)$ and $(0,\pm a)$, where $%
a=0.05\xi $ and $\xi =\hbar \upsilon _F/\Delta _0$ is the coherence length
of {\it d}SC state, two resonant peaks appear below and above the Fermi
level, respectively. However, on the next nearest neighbor sites ${\bf r}%
_0=\pm (a,a)$, the LDOS obtained is very much similar to that on the Zn
impurity site. All these features are in good agreement with the STM
experimental results \cite{pan-nature}.

Finally, the spatial distribution of the LDOS of the {\it d}SC
quasiparticles is  calculated. The LDOS around the Zn impurity at the main
resonance energy is displayed in Fig.8a\ for the hole-like excitations $%
\omega =-\Omega _{res}$ and in Fig.8b for $\omega =\Omega _{res}$ for the
electron-like excitations as a function of spatial variables for $c=1,2,$
and $3$ with $\epsilon _d/\Delta _0=-0.05$ in a logarithmic intensity scale.
The quasiparticle resonances induced by the doped non-magnetic impurity are
extended from the impurity to one coherence length of {\it d}SC. The largest
amplitude of the quasiparticle resonance occurs at the neighborhood of the
impurity, particularly at the next nearest neighbors, and the local
electronic structures at distances $r\ll \xi $ drastically differ in Fig.8a
from Fig.8b. However, further away from the impurity site, both structures
of the LDOS are almost the same, where the enhanced LDOS is extended along
the gap maxima directions. Moreover, the spatial LDOS of the {\it d}SC
quasiparticles is not enhanced as the strength of the scalar potential is
increasing. The LDOS pattern in the left column for $c=1$ case is closer to
the distribution of LDOS obtained from the STM\ spectra \cite{pan-nature}.
These results seem to show that the strength of potential scattering in the
present model Hamiltonian, needed to describe the  experimental data is not
as strong as assumed in previous studies with purely potential scattering 
\cite{balatsky,flatte}.

\section{Conclusions}

In this paper, we propose an effective model Hamiltonian of an Anderson
localized electron (hole) hybridizing with $d_{x^2-y^2}$-wave BCS type
superconducting quasiparticles supplemented with an attractive scalar
potential at the impurity site. Within a unified framework, we can describe
the non-magnetic (Zn) and magnetic (Ni) impurity scatterings for cuprate
superconducting quasiparticles of optimally doped high T$_c$ materials.

Due to the nature of $d$-wave superconducting quasiparticles, both quantum
scattering on localized electron and scalar potential scatterings are
relevant interactions in the low energy limit, showing a strong interplay
with each other and leading to localized resonant states below the maximal
superconducting gap. In the Ni case, two localized resonant states are
formed above the Fermi level in the LDOS at the doped impurity site, while
in the Zn case a sharp resonant peak caused by the localized hole is found
just below the Fermi level to dominate the LDOS at the Zn impurity site,
accompanied by a small and broad bound state above the Fermi level induced
by the potential scattering. These characteristic features are in exact
agreement with experiments \cite{pan-nature,hudson-nature}. Moreover, the
spatial patterns of the resonant states in Zn and Ni doped cases are quite
different in STM experiments. It is significant that these differences can
be reproduced easily in our simple model. {\it The key assumption is that
the $3d$ electron is located on the Ni site, while it is distributed over
nearest neighbor Cu ions in the Zn-doped case due to the local
antiferromagnetic background.} The excellent agreement with experiments
convincingly demonstrates the correctness of this assumption and the
self-consistency of our model. We would emphasize these are generic features
of our model, not due to adjustable parameters in the theory to fit the
experiments.

We should point out that in both cases, there are no Kondo screening effects
on the local magnetic moment due to the localized electrons, {\it i.e}., the
local magnetic moments do not couple to the superconducting quasiparticles
in the low energy limit, and the dominant behavior is the single hole
occupied state, characterized by a strong hybridization between the
localized and delocalized electrons. Moreover, from the calculations of the
spin relaxation functions for the localized electron in both Ni and Zn doped
materials, we find that the additional $3d$ localized hole displays a weak
magnetic oscillation. Such a magnetic behavior is consistent with the
spin-1/2 magnetic moment observed in NMR measurements.

We would like to mention that our results do not contradict those of the
Kondo model study for systems with reduced density of states at the Fermi
level \cite{sachdev,ting,vojta,ingersent,fujimoto,vojta2} or $d$-wave
superconductors. The important factor in those studies is the presence of a
strong particle-hole asymmetry, due to, for example, the potential
scattering. The behavior of systems controlled by the ``strong asymmetric
coupling'' fixed point is very similar to what we have obtained here. In
particular, there is no single Kondo resonance peak at the Fermi level in
that case, either. Instead, the resonance is shifted from the Fermi level by
the energy of the order $T_K$. In our opinion, identifying this resonance
peak to the standard Kondo screening effect is somewhat misleading. The NMR
experiments in Ni, Zn-doped cuprates samples \cite{nmr1,nmr2,bobroff} can be
interpreted using the ``Kondo temperature'' $T_K$ as a characteristic energy
scale. Recent experiment \cite{bobroff} seems to show such a scale exists
even in the superconducting state. On the contrary, the SQUID measurements
do not exhibit any signature of such a ''Kondo'' scale.\cite{nmr3} We
believe the resistivity measurements would be able to identify contributions
from the Kondo scattering. However, our paper and related studies \cite
{sachdev,ting,vojta,ingersent,fujimoto,vojta2} demonstrate that in low
temperature superconducting state one should not expect the standard Fermi
liquid behavior, corresponding to the Kondo strong-coupling limit. To our
feeling, the effective Anderson model used in our paper has the advantage of
treating different physical situations in a unified self-consistent fashion.

A few related remarks are in order: \newline
(1) In this paper we did not address the issue of the magnetic moment
formation in non-magnetic doping case. A lot of efforts have been devoted to
this problem \cite{see}, but it is still not fully understood, in
particular, the issue of spinon confinement which is the key argument for
the "proof", is entirely open. What we have been doing is to assume the
existence of the local moment, described by the Anderson model and find out
that the final results are consistent with the starting assumption. \newline
(2) To reconcile the discrepancy of the potential scattering model \cite
{balatsky} with the STM experiments on Zn-doped BSCO, namely, the presence
of a strong resonance peak at the Zn site in experiments, the argument of
tunneling matrix element between layers (so-called ``filter'' effect) \cite
{martin,ting2} was suggested. We have not included this effect in our
calculations to confront the STM experiments. To our knowledge, this issue
is still open, and the Kondo model calculation \cite{sachdev,vojta} seems to
show the same --- irrelevance of this effect.\newline
(3) We have not considered the effect of disorder on the quasiparticle
density of states in $d$SC which may not vanish at the Fermi level. This is
still an issue under debate.\cite{atkinson} Our hope is that the main
conclusions of our paper will survive in the limit of dilute doping. The
case of strong disorder is beyond the scope of our present work.\newline
(4)The interplay of magnetic and non-magnetic scatterings is always
important for superconducting states even in the $s$-wave case \cite{yulu}.
It is true that the potential scattering alone cannot give rise to a bound
state inside the gap in that case (Anderson theorem). However, the presence
of the potential scattering will affect the position of the bound state due
to the magnetic moment scattering (see first paper of Ref. \cite{yulu}).
What we have shown is that the scattering behavior in the $d$-wave nodal
direction is quite different from the $s$-wave superconductor, whereas in
the gap maximum direction it is rather similar. For the low-energy physics,
the quantum magnetic scattering is the dominant effect. Finally, we would
emphasize once again that our result is valid only in the strong coupling,
large $U$-limit of the Anderson impurity in a $d$-wave superconductor, which
is very different from that for the conventional normal metallic state.

Acknowledgments

We would like to thank K. Ingersent, S. H. Pan, M. Vojta and T. Xiang for
useful discussions. One of the authors (G.-M. Zhang) would like to thank the
hospitality of Abdus Salam International Center for Theoretical Physics,
where part of this work is completed, and acknowledges the financial support
from NSF-China (Grant No. 10074036 and 10125418) and the Special Fund for
Major State Basic Research Projects of China (G2000067107).

\newpage

\begin{center}
{\bf Figures Captions}
\end{center}

Fig. 1. The DOS $N_d\left( \omega \right) $ of the localized electron, (a)
for $c=1$ and (b) for $c=2$, with $\epsilon _d/\Delta _0=$ $-0.05$, $0.05$, $%
0.15$, and $0.30$, denoted by solid, dashed, dotted, and dash-dotted lines,
respectively.

Fig. 2. The imaginary part of the self-energy for the localized electron is
delineated in (a), and the corresponding local spin relaxation function $%
S_d(\omega )$ in (b), for $\epsilon _d/\Delta _0=$ $-0.05$ and for $c=0$, $1$
and $2$, represented by solid, dashed, and dash-dotted lines, respectively..

Fig. 3. The LDOS $N_c\left( r{\bf =}0.07\xi ,\omega \right) $ of the {\it d}%
SC quasiparticles for different potential scattering strengths $c=2$ and $5$%
, with $\epsilon _d/\Delta _0=$ $0.0$ (solid line), and $0.3$ (dashed lines)
in units of $N_F$. (a) and (c)\ are along the directions of the gap maxima,
while (b) and (d) along the directions of the gap nodes.

Fig. 4. The spatial distributions of the conduction electron LDOS around the
Ni impurity at the first resonant energy for $c=1,2$ and $3$. (a)\ $\omega
=-\Omega _{res}^1$ and (b) $\omega =+\Omega _{res}^1$. Here $\epsilon
_d/\Delta _0=-0.05$, and a logarithmic intensity scale is used. The ranges
of $x$ and $y$ in each subplot are set from $-0.6\xi $ to $+0.6\xi $, and $%
\xi $ is the coherence length.

Fig. 5. The DOS of the localized electron $N_d\left( \omega \right) $. (a)
corresponds to $c=2$, and (b) to $c=5$, with $\epsilon _d/\Delta _0=$ $-0.4$%
, $-0.2$, and $0.0$, denoted by solid, dashed, and dash-dotted lines,
respectively.

Fig. 6. The imaginary part of the self-energy for the localized electron is
displayed in (a), and the corresponding spin relaxation function $S_d(\omega
)$ in (b) for $\epsilon _d/\Delta _0=$ $-0.2$ and for $c=1,2$ and $5$,
denoted by solid, dashed, and dash-dotted lines, respectively.

Fig. 7. The LDOS of the {\it d}SC quasiparticles on the Zn impurity site, on
the nearest neighbor sites, and on the next nearest neighbor sites with $%
\epsilon _d/\Delta _0=$ $-0.3$ (solid line), and $-0.1$ (dashed lines) in
units of $N_F$. (a) corresponds to $c=2$, and (b) to $c=4$.

Fig. 8. The spatial distributions of the LDOS of {\it d}SC quasiparticles
around the Zn impurity at the resonant energy for $c=1,2$ and $3$. (a)\ $%
\omega =-\Omega _{res}$ and (b) $\omega =+\Omega _{res}$. Here $\epsilon
_d/\Delta _0=-0.2$ and a logarithmic intensity scale is used. The ranges of $%
x$ and $y$ in each subplot are set from $-0.3$ to $+0.3$ in units of the
coherent length $\xi $.

\end{document}